# APPLICATION FIELDS OF
# HIGH-TEMPERATURE SUPERCONDUCTORS

**Roland HOTT**

Forschungszentrum Karlsruhe, Institut für Festkörperphysik,
P. O. Box 3640, 76021 Karlsruhe, GERMANY

Email: roland.hott@ifp.fzk.de

## 1. Introduction

For classical superconductors it took about half a century from their discovery to arrive at technically applicable materials. Only about a third of this time has elapsed for High-Temperature Superconductors ("HTS") where a number of technically applicable materials species is now already available.

Epitaxial HTS *thin films* achieve excellent superconducting properties (critical temperature $T_c > 90$ K; critical current density $J_c$ (77 K, 0 T) $> 10^6$ A/cm$^2$; microwave surface resistance $R_s$(77 K, $f = 10$ GHz) $< 500$ μΩ, $R_s$(T, $f$) $\propto f^2$ ) that are well-suited for superconductive electronics. They are already in use in commercial and military microwave filter systems. HTS *Josephson junctions* are available which can be used for the construction of highly sensitive magnetic field sensors ("Superconducting QUantum Interference Devices", "SQUIDs") and are also tested for active electronic devices that may broaden the range of HTS thin-film applications. *Melt-textured* HTS *bulk material* shows superb magnetic pinning properties and may be used as high-field permanent magnets. In spite of the ceramic nature of the cuprate oxides, flexible HTS *wire* or *tape conductor* material is obtained either by embedding HTS as thin filaments in a silver matrix or by coating of metal carrier tapes.

Among these technical HTS materials, Ag-sheathed Bi-HTS conductors represent the only exception to the rule that strong biaxial texture is necessary to achieve technically meaningful currents. A main reason is here the good mechanical and electrical contact with the Ag matrix which allows high current flow under the relaxed condition that a little detour via Ag is possible in case the direct current transfer between the HTS grains is blocked. These arguments apply as well to "Ag-impregnated" Bi-2212 bulk which is commercially available in sizes of several 10 cm with homogeneous $J_c$(77 K, 0 T) of several kA/cm$^2$ [1]. Much higher critical current densities are achieved in polycrystalline HTS materials, but their use implies a higher risk for applications: In all these cases, substantial $J_c$ reduction compared to the $J_c$ (77 K, 0 T) ~ 1 MA/cm$^2$ encountered in well-textured material with a sufficient amount of pinning centers always indicates a high degree of structural and electrical inhomogeneity with potentially devastating consequences: Forced current flow through defective materials regions may lead to a local quenching of superconductivity and the creation of "hot spots" [2]. The deposited quench energy leads then again to a structural and chemical modification of the HTS material in the neighborhood of these "hot spots". This enlarges the quench zone and may finally end up in the destruction of the superconductor. This quench mechanism is different from the situation in classical superconductors where the quenching is caused by insufficient heat transfer due to the freezing of the phonon mechanism at LHe temperature. At the much higher operating temperature of HTS this is not such a critical issue since the heat distribution by phonon is here still very effective [3].

Cooling is the main concern for a future breakthrough of HTS applications. The market acceptance for HTS-based systems will depend critically on the availability of reliable and inexpensive cooling systems that are "invisibly" integrated in these systems.

## 2. Electronics

The low microwave losses of HTS thin films [4] enables the coupling of an unprecedentedly large number of resonators to *microwave filter* devices with much sharper frequency characteristics than conventional compact filters [5, 6].

In the US, the military interest in HTS filters in aircraft electronics, e.g. for better rejection of interference noise in aircraft radar systems, is still strong. In US mobile phone communication systems, HTS microwave filter subsystems are already a commercially available solution for problematic radio reception situations [7]: Improved noise data compared to conventional solutions result in a lower percentage of dropped calls; in rural areas rf coverage can be achieved by a smaller number of base stations; the use of HTS filters could allow to reduce the rf power of the handsets in urban areas. The miniaturization potential of HTS filters should be most appreciated in communication satellites [8], however, respective projects were not very fortunate [9, 10].

In *Josephson samplers*, the fast voltage reaction of a Josephson junction on a signal current that exceeds the critical current allows the determination of (repeated) signal forms with ps resolution. A first commercial system fabricated in Niobium technology disappeared again from the market soon after its introduction in 1987 due to problems with the complexity of the LHe cooling and high maintenance cost (as well as the introduction of semiconductor samplers with comparable bandwidth). An HTS Josephson sampler system under development at NEC has already demonstrated the resolution of a 5.9 Gbps digital waveform [11].

The periodicity of the electric characteristics of superconducting loops as a function of the introduced magnetic flux can be used for the construction of superconducting *AD converters*. Flux quantization can help to implement feedback loops with quantum accuracy [12]. First test of components [14,15,16] and complete devices [13] based on ~ 10 HTS Josephson contacts demonstrated technical feasibility.

Recent simulations of the noise-induced bit error rate in HTS *digital circuits* indicate good chances even for $LN_2$ temperature operability [17] for suitable circuit technology [18] since connecting Josephson junctions in a circuit results in improved energetic separation of the different switching states as compared to the isolated junction behavior [19].

An HTS programmable *voltage standard* based on an array of 136 YBCO bicrystal junctions operated at 64 K could be synchronized at 25 - 40 GHz showing stable voltage steps up to 9 mV with an accuracy better than $10^7$ [20].

Among all the tested HTS *transistor* concepts [21] only Vortex-Flow Transistors (VFT) have demonstrated large gain-bandwidth potential [22]. Quasiparticle Injection devices [23] or Superconducting Field Effect Transistors ("SuFETs") [21,24,25,26,27] are still far from technical applications.

## 3. Sensors

*Superconducting QUantum Interference Devices* ("*SQUIDs*") are superconducting loops with integrated Josephson contacts which can be used as the most sensitive magnetic field sensors. The magnetic field resolution of HTS *SQUIDs* [28] operated at $LN_2$ temperature has arrived at values comparable to commercial LHe-cooled LTS SQUIDs, only one order of magnitude above the record values of LTS SQUIDs at 4 K [29]. However, a large commercial impact is only expected for HTS SQUID systems that are able to observe magnetic signals even in the presence of disturbing background fields without the burden of magnetic shielding [30,31,32,33,34]. Mobile non-shielded HTS SQUIDs were tested with success for biomedical [31,35], non-destructive evaluation [36,37] and geophysical applications [38].

In superconducting *bolometers*, the sharp superconducting transition as a function of temperature is used as very sensitive thermometer which allows to measure with high sensitivity the heating of a thermally connected absorber under electromagnetic irradiation. The high quantum efficiency of semiconductor detectors restricts yet the application interest to the infrared wavelength region $\lambda > 12$ μm where no adequate semiconductor material is available [39,40,41,42]. As $LN_2$ cooling is common practice for semiconductor IR radiation sensors, HTS bolometers could easily be inserted in such detector systems. A Gd-123 based bolometer has met the specifications for a sensor for IR observation of OH molecules in the atmosphere in a satellite project [39]. Hot-electron bolometers with small thermal relaxation time are tested as *mixers* for high-frequency signals in the range of several 100 GHz [43].

The interaction of an electromagnetic wave in a Josephson contact with the Josephson oscillations results in an analog way of doing a *Hilbert transformation of the radiation spectrum* where the Hilbert transform can be derived from the IV characteristics [44,45]. Experiments based on a YBCO bicrystal contact demonstrated that a spectral resolution of ~ 4 GHz can be achieved in a bandwidth from ~100 GHz to 3 THz [44]. The device has been tested with success at DESY for the determination of the bunch length of a pulsed electron beam by means of the emitted rf spectrum.

## 4. Magnets

The first commercial impact of HTS on superconducting magnet systems are *current leads* that substantially reduce the heat load to the cold magnet system. Whereas for normal metals the Wiedemann-Franz relation between electrical and thermal conduction results in a universal minimum value ~ 1 W/kA for charge transport to a 4 K heat load, HTS in the superconducting state offer a large current capability at the low thermal conductivity of oxide ceramics [46]. Bi-2212 bulk based leads demonstrated a reduction of the 4 K heat load by a factor ~ 5 compared to conventional leads in dc as well as in ac application [47,48]. Current leads based on Bi-2223 tapes with AgAu alloy sheath (with one order of magnitude lower thermal conductivity compared to Ag) achieve similar performance [49]. The intrinsic parallel shunt provided by the normal metal sheet and the reduced mechanical sensitivity are attractive features with respect to the emergency situation in case of a quench of the connected LTS coil. For big LTS research magnets the heat load due to the current leads amounts to 30 - 90 % of the total 4 K heat load. At present, the largest order for HTS current leads is expected from CERN where 64×13 kA, 310×6 kA and 750×600 A leads will be required for the Large Hadron Collider project [50]. Several types of HTS hybrid current leads have been tested [50]. At Fermilab 30 - 40 of the existing 50 pairs of 5 kA current leads are planned to be substituted by HTS hybrids. The gain of 30 % in heat load to the LHe bath will be used for lower operation temperature of the LTS magnets which will then allow stronger particle acceleration. HTS current leads for much higher currents in future fusion magnets (40 - 60 kA) are under development [51]. The reduction of the heat load by means of HTS current leads and the availability of efficient cryocoolers enabled "cryogen-free" LTS magnet systems which can be operated by means of conduction cooling without the requirement of LHe coolant [52].

The extraordinarily high critical magnetic fields of HTS (estimated: ~ 100 T) allow the flow of supercurrents in much higher magnetic background fields than in classical superconductors. Thus HTS magnets are able to generate an additional magnetic field even in the environment of a strong magnetic background field [53]. The critical issue is the mechanical stress in the HTS conductor as a consequence of the magnetic field. A Bi-2212 magnet successfully generated 5.42 T in a backup field of 18 T provided by a classical superconducting coil system [54]: The total magnetic field of the superconducting magnet system reached 23.42 T, which is the highest magnetic field ever achieved by a super-conducting magnet. The goal of such activities are high field magnets for Nuclear Magnetic Resonance (NMR) experiments where the persistent mode operation of the superconducting coils is used to fulfill the extreme requirements of temporal stability. The respective time constants for the decay of the magnetic field in present HTS coils are still about an order of magnitude below these requirements.

The higher operation temperature of HTS magnets is of interest for many applications where classical superconducting coils can not be used due to strong heat generation or heat leakage to the cold part [55,56]. In a US project, DuPont assembled, installed and tested a 0.2 m bore 3 T HTS magnet for minerals and chemical separation[1] [57]. A full-scale pre-production unit with 0.5 m warm bore 3 T HTS magnet will be the next step in this project. In another US project, Oxford Superconducting Technology plans to build a cost-effective, open-geometry MRI system based on a HTS 0.2 T magnet [57].

---

[1] In a Japanese project, a mobile water purification system based on magnetic separation by strong melt-textured YBCO bulk cryomagnets is under construction [58].

*Superconducting magnetic bearings* ("SMB") composed of melt-textured YBCO pellets and permanent magnets achieved a coefficient of friction (drag-to-lift ratio) of $\sim 10^{-9}$ [59,60,61]. They are investigated for the levitation of rotors in flywheels [62] and motors [63]. Compared to active magnetic bearings based on permanent magnets and active rotor control, SMB can offer similar levitation forces [64], similar power consumption (cooling power / control power) and lower friction, but unfortunately only much lower stiffness [65]. In China, a first man-loading HTS Maglev test vehicle was tested successfully with up to five people and a total weight is 530 kg at a net levitation gap > 20 mm on a 15.5 m long guideway consisting of two parallel permanent magnetic tracks [66].

## 5. Power Applications

Superconductors can help to increase the efficiency of components for power applications by means of their lower losses as well as to reduce volume and weight by means of their potential for high power density [67]. However, reliability of these large scale devices is here of even higher importance than for other potential application fields. HTS demonstrators are still on their way of trying to convince power companies about their trustworthiness even over a long lifetime.

*Power cables* are in general about a factor of 10 more expensive than overhead lines. This restricts their application to urban areas. Experiments with LHe-cooled LTS cables in the 70's were technically successful [67], but their introduction into power grids turned out to be economically viable only for power transfer > 5 - 10 GVA. For $LN_2$-cooled HTS cables, this break-even point may be reduced to $\sim$ 300 - 500 MVA, if the cables can be retrofitted to already existing cable ducts. This would allow for an increase of power transfer by a factor $\sim 3$ at a reduction of the voltage from $\sim$ 400 kV to $\sim$ 100 kV [68]. Therefore, such a retrofit is especially attractive for metropolitan areas with still growing power demand, such as Tokyo, where no more additional cables can be inserted into the cable ducts, and voltage has been increased right up to the highest justifiable level. Besides this civilian application, the US Navy is also interested in lightweight small-sized cables for battleships [69]. $LN_2$ cooling offers distinct advantages compared to conventional oil-cooling with respect to pollution in case of leakage and with respect to the risk of fire. The cable market amounts to 5 - 10 % of the total power equipment market. The low operational self-fields of $\sim 0.1$ T of power cables are within the acceptable limit of present commercial Bi-2223/Ag tapes at $LN_2$-cooled operation. In a Danish project, a 30 m 3-phase HTS cable with a voltage rating of 30 kV and a power rating of 104 MVA has been installed and tested under realistic conditions in a substation in the electric grid of Copenhagen [70]. In a Japanese cable project, a 100 m 3-core 66 kV/1 kA/114 MVA has been tested successfully for one year duration [71]. In a US cable project, Southwire installed a 30-m 3-phase cold dielectric cable system (12.5 kV, 1250 $A_{rms}$, 27 MVA) which is still in test operation since February 2000 [57]. In another US cable project, a 3-phase 120 m cable system has been installed in a power station [72] but severe problems with damage of the flexible cryostat of the warm-dielectric type cable have prevented until now operation of 2 of the 3 phases. Recently, two 300 m HTS cable projects have been approved in the US which are planned to be installed in electricity distribution systems in 2005 [73].

The *fault current limiter* ("FCL") is unanimously addressed by utilities as a very attractive component, integratable in existing power grids [74]. It prevents overloads from the grid components thus enabling longer lifetimes and avoiding investment cost due to the usually practiced overdimensioning. For FCLs rated at a nominal power of at least 10 - 20 MW as smallest practical size, an annual market volume of ~ 1 billion US-$ has been estimated. The current limitation is based on the quench of a superconductor due to a current exceeding its critical current, resulting in a very rapid tremendous increase of its electrical resistance. The material problem to be solved is the spatial homogeneity of the local critical current density on the length scale of the thermal diffusion length because otherwise the power is not dissipated homogeneously, but directed into hot spots, destroying the material. This homogeneity has been achieved up to now only for Bi-2212 bulk [75], Bi-2223 tapes [79] and YBCO tapes [80] and YBCO thin films [81]. A 1.2 MW-3 phase-FCL by ABB based on Bi-2212 bulk rings acting as secondary winding of a transformer-like device has been tested for one year in a Swiss power plant without major problems [75]. In a test series, a prospective fault current of 60 kA was limited to 700 A within the first 50 Hz-half-wave. The time for recovery to the full operational state after a quench amounted to a few seconds. Meanwhile, ABB presented a single phase 6.4 MVA device with a different FCL design based on 30×40 cm$^2$ sized Bi-2212 sheets structured into long length meanders [76]. In a German collaboration of industrial companies, research institutes and power utilities, a 15 MVA FCL demonstrator of the resistive type for the 10 kV voltage level based on Bi-2212 spirals is planned to be assembled and field-tested this year [77]. The limiter is designed for coupling two 125 MVA utility grids.

HTS *transformers* can be built at smaller volume and weight compared to their conventional counterparts [78]. With respect to the risk of fire and environmental aspects, LN$_2$ is a much more pleasant coolant than oil used today. However, the main advantage of SC transformers is their potential for an operation at a power level exceeding the rated power by up to 100 % even over a period of several hours: This exceptional operation mode merely requires more cooling power, but does not lead to increased wear as in the case of conventional transformers where even ~ 10 % overload causes thermal damage in the insulation. With regard to energy losses, a quarter of the losses of 5 - 10 % of the transported power in power distribution systems are due to transformers. The use of a winding based on a HTS conductor with fault current limiting functionality [79] is an additional option. ABB tested a 630 kW - 3 phase transformer based on Bi-2223/Ag tape conductor successfully at a Swiss utility for one year under regular operational conditions [82]. In a US project, a 5/10 MVA prototype based on Bi-2212 surface-coated conductor is under development [83]. Siemens developed a 1 MVA demonstrator transformer for railway applications which has been tested with success [84].

For large synchronous *motors*, 50 % volume and loss reduction compared to a conventional motor is expected [85, 86]. For the US Navy, this is a very attractive feature since their new generation of surface ships is based on electric drives due to lower operating and support cost [69]. The respective project for has meanwhile arrived at a 3.7 MW engine which was tested up to a peak power of 5.2 MW [87]. Furthermore, the rotor of a 5 MW engine has been completed which will now be assembled until mid of this year for the first HTS ship propulsion motor. The US Navy is also interested in a super-quiet homopolar engine with regard to improved stealth [69,88]. Siemens built a synchronous machine consisting of an HTS rotor and an air core stator that achieved a maximum continuous power of 450 kW and a short term maximum power of 590 kW at 1500 rpm [89]. Rotor cooling is provided by a GM "off the shelf" cryocooler, the typical requirement being about 30W @ 25K. The output power was limited by stator cooling. The machine was tested under different operational modes, including motor mode as well as generator mode. Siemens started now work on a 4 MW ship propulsion motor. Motors based on the strong magnetizability of melt-textured YBCO bulk material with an output power of 1-37 kW and current frequencies 50 and 400 Hz have demonstrated in $LN_2$ operation a 4–5 times better specific output power per weight than conventional electric machines [90]. Such motors could find an application in cryopumps for cryogenic liquids which could be used simultaneously as coolant for the HTS material.

HTS *generators* are of considerable military interest in the US: The US Navy is interested in lightweight small-sized HTS power systems (generators, cables, fault-current limiters, transformers, motors) for battleships [69]. The US Air Force develops a new generation of aircrafts where the hydraulics will be completely replaced by electronics. HTS generators would be extremely desired provided they are available within the next 3-5 years [91]. In a civilian US project, General Electric is developing a 100 MVA HTS power generator for commercial entry applications [57]. It is based on a previously constructed 15 MVA model.

Superconductivity offers two extremely different ways for *energy storage* [92]: In *Superconductive Magnetic Energy Storage* ("*SMES*") systems, a SC coil stores magnetic field energy. The energy can be transferred very rapidly via power electronics to or from the grid [93]. SMES systems based on classical superconductors are already in commercial use for the improvement of the power quality. HTS SMES systems are of interest with respect to size and volume reduction [94]. In *flywheels* based on superconductive magnetic bearings [95], electric power is transformed to kinetic energy via a motor-generator combination controlled by power electronics. Due to the mechanics involved, the energy can not be transferred as fast as by SMES. However, the density of storable energy can be much higher [92] and operational losses are expected to be significantly lower [96]. In a US project, a 10 kWh / 3 kW system has been assembled and installed [57]. In a Japanese project, a flywheel demonstrated a maximum energy storage of 1.4 kW h at 20 000 rpm [97]. A large-size bearing for a 25 kW h flywheel has been tested. In Germany, a 10 kWh / 1-2 MW system is under construction which is planned to be field-tested next year [98].

## 6. Cryocoolers

The most visible aspect of superconductivity for any new potential customer of superconductive devices is not superconductivity but the required cryocooling. As for the cost, as a rule of thumb about 10 % of the total system prize seems to be an accepted level for most of the applications. The goal of a US project for low price cryocoolers (~ 1000 US $ at annual sales of ~ 10 000 coolers) which should reliably provide a cooling power of some W at 50 - 150 K over a period of 3 years [99] has not been achieved yet[2]. The larger cooling requirements of HTS power applications can already be met by reliable commercially available cryocoolers, but, e.g. for transformers, only at a price which amounts to sales price of a conventional transformer. This economic mismatch may only be overcome by much higher sales figures

---

[2] An excellent topical survey on low-power cryocoolers can be found in [100].

and technical simplification of the cryocoolers. The technical boundary conditions for the selection of the cryocooler are minimal disturbance of the cooled device, e.g. by mechanical vibrations, and good cooling efficiency expressed by the ratio of cooling power at a certain operating temperature divided by the required electric power. In commercial US mobile phone microwave filter systems, cryocoolers with a cooling power of several W in the $LN_2$ temperature region have already demonstrated their utilizability in more than 20 million cumulative hours of operation in the field from which a Mean Time Before Failure ("MTBF") of $\sim 800\,000$ hours has been estimated [101].

Different cooling principles are available for cooling to the $LN_2$ or even LHe temperature region. *Joule-Thomson* ("JT") coolers are based on the heat exchange during the expansion of a continuous gas flow. A commercial JT cooler with a cooling power of 3 W @ 77 K has been used successfully for HTS SQUID cooling [102].

*Stirling*, *Gifford-McMahon* ("GM") and *Pulse Tube* ("PT") coolers are all based on an oscillating gas flow and a regenerative heat exchange. The heat extraction depends on the oscillating frequency f of the gas flow and the phase angle between pressure p and volume V during an oscillation period [103,104]. In an analog electric network description these coolers can be regarded as RC phase shifters. In Stirling and GM coolers a piston is moving in the cold part in which the regenerator is usually already integrated. The PT cooler can be conceived as a GM or Stirling cooler where the gas flow is steered in such a way as if a "gas piston" was moved according to the particular cooler operation mode.

*Stirling* coolers are usually operated at frequencies f $\sim 20$ - 60 Hz and achieve minimal temperatures $T_{min}$ = 30 - 55 K. The weight of coolers commercially available at 3000 to 20 000 US $ reaches from 300 g (150 mW @ 80 K, $P_{el}$ = 3 W) to 35 kg (18 W @ 80 K, $P_{el}$ = 375 W). The prize extremes are given by a 250 000 US $ cooler manufactured for the US HTS Space Experiment [9] (specified MTBF: 45 000 h) on one side and a cryocooler used in the commercial US mobile phone microwave filter systems with an estimated prize < 2000 US $ for annual sales numbers > 1000. For the US military aircraft HTS application plans, available Stirling coolers seem to satisfy already the specifications concerning efficiency, reliability and cost. A disadvantage of these coolers especially for SQUID applications is the small distance from the compressor to the cold head of at most 40 cm.

Coolers of the *Gifford-McMahon* type are sold worldwide in numbers of $\sim 10\,000$ per year for cryopumps to the electronics industry [105] and for the assistance of the LHe cooling of LTS MRI magnets by means of 20 K heat shields. Compressor and cold head are usually separated units. The pressure oscillations are produced by a rotary valve in the cold head. Due to this construction principle GM coolers are on one side substantially heavier than Stirling coolers with the same cooling power and achieve only $\sim 1/3$ of their cooling efficiency. On the other side, they are more robust than Stirling coolers. This helped the GM coolers to establish themselves as standard coolers for temperatures < 30 K providing a cooling power of 0.5 - 3 W @ 4.2 K at an input power of 1 - 12 kW [106]. The suggested service intervals are 5 years for the compressor (with an annual oil exchange) and 1 - 3 years for the rotary valve and the seals. 4 K GM coolers are today available at a price of $\sim 40\,000$ US $ which is estimated to be reducible to $\sim 5000$ US $ at drastically increased sales numbers.

*Pulse Tube* coolers have no moving parts in the cold region. Therefore no expensive high-precision seals are required and the cold head can be operated without any service inspection. 1-stage PT coolers achieve cooling powers of several W in the $LN_2$ temperature region (up to 166W @ 80 K at $P_{el}$ = 3.42 kW [107]). With 2-stage PT coolers [108] 0.5 W @ 4.2 K at $P_{el}$ = 6.3 kW and 0.17 W @ 4.2 K at $P_{el}$ = 1.7 kW have been demonstrated. Already in 1993 a PT with 0.8 W @ 80 K at $P_{el}$ = 33 W and a weight of 2 kg achieved the performance of a comparable Stirling cooler [109]. Meanwhile, Stirling-type PT (high operation frequency: $f \sim 50$ Hz) have arrived at an even better efficiency at 77 K than their conventional counterparts, GM-type PT (low operation frequency: $f \sim 5$ Hz) show at 4 K comparable efficiency [110]. Stirling-type PT with a cooling power of 12 W @ 80 K are commercially available where only the compressor requires servicing every 20 000 h. Miniature PT are in use for 10 year long-life space applications [109]. PT coolers were used with success for the cooling of HTS SQUID and rf devices [102, 111]. A completely non-metallic PT cooler has been tested with success for low magnetic noise SQUID cooling [111].

## 7. Conclusion

Recalling the situation in the late 1980 in the "prime time" of High-Temperature Superconductors immediately after their discovery, it is amazing to see which tremendous progress has been achieved in the meantime: HTS filters in mobile phone base stations, HTS SQUID operation in regular technical environment, HTS magnets generating a 5 T field in addition to a 18 T background, 100 m long HTS power cables are more than what could be expected on realistic grounds in such a short period of time. Steady improvement of the HTS materials basis will surely widen this spectrum of applications within near future.

## Acknowledgement

I would like to thank T. Habisreuther, P. Komarek, M. Lakner, T. Scherer, C.W. Schneider, P. Seidel, M. Siegel, H. Töpfer and G. Thummes for critically reading the manuscript and for helpful comments.